%% file: main.tex
\documentclass[conference,compsoc]{IEEEtran}

\usepackage[utf8]{inputenc}
\usepackage{amsmath, amssymb}
\usepackage{graphicx}
\usepackage{hyperref}
\usepackage{url}
\usepackage{cite}
\usepackage{tikz}
\usepackage{booktabs}
\usepackage{multirow}
\usepackage{tabularx}

\usetikzlibrary{positioning, shapes.geometric, arrows}
\usepackage[linesnumbered,ruled,vlined]{algorithm2e}  % For algorithm block
\SetKwInput{KwIn}{Input}      % Customize input
\SetKwInput{KwOut}{Output}    % Customize output

\newcommand{\etal}{\emph{et al.}\xspace}
\newcommand{\method}{MIA-EPT}

\usepackage{tikz}
\usetikzlibrary{positioning, arrows.meta}  % For node positioning and arrow styles

\usepackage{amsmath}    % For math formatting
\usepackage{amssymb}    % For math symbols

\title{\method: Membership Inference Attack via Error Prediction for Tabular Data}

\author{
Eyal German\textsuperscript{*}, Daniel Samira\textsuperscript{*}, Yuval Elovici, Asaf Shabtai\\
Department of Software and Information Systems Engineering, Ben-Gurion University of the Negev \\
\{germane, samirada\}@post.bgu.ac.il, \{elovici,shabtaia\}@bgu.ac.il \\
\textsuperscript{*}Equal contribution
}

\begin{document}
\maketitle

% \begin{abstract}
% Synthetic data generation is a promising technique for privacy-preserving data sharing. Diffusion models for tabular data achieve high fidelity but may inadvertently leak information about training records. Membership Inference Attacks (MIAs) aim to detect whether a particular record was used to train a model, posing privacy risks. We propose MIA-EPT, a novel black-box MIA framework that leverages prediction error patterns from helper models trained on synthetic data. MIA-EPT was evaluated in the MIDST 2025 challenge, where it achieved 2nd place in the Black-box Multi-table track, revealing significant privacy leakage in state-of-the-art tabular diffusion models. Our full code can be found at \url{https://github.com/eyalgerman/MIA-EPT}
% \end{abstract}

\input{content/abstract}

\input{content/introduction}

\input{content/related_work}

\input{content/method}

\input{content/evaluation}

\input{content/results}

\input{content/conclusion}

\bibliographystyle{IEEEtran}
\bibliography{sample}

\input{content/appendix}

\end{document}

%% file: content/abstract.tex
\begin{abstract}
% Synthetic data generation plays an important role in enabling data sharing, particularly in sensitive domains like healthcare and finance. 
% Recent advances in diffusion models have made it possible to generate realistic, high-quality tabular data.
% However, these models can inadvertently memorize individual training records during training and leak sensitive information through the synthetic data they generate, raising significant privacy concerns.
% Membership inference attacks (MIAs) exploit such vulnerabilities by attempting to determine whether a specific data record was part of a model’s training set. 
% While MIAs have been studied extensively in the context of image and text data, their application to generative models trained on tabular data remains underexplored, despite the distinct privacy risks posed by structured attributes and limited record diversity.
Synthetic data generation plays an important role in enabling data sharing, particularly in sensitive domains like healthcare and finance. Recent advances in diffusion models have made it possible to generate realistic, high-quality tabular data, but they may also memorize training records and leak sensitive information. Membership inference attacks (MIAs) exploit this vulnerability by determining whether a record was used in training. While MIAs have been studied in images and text, their use against tabular diffusion models remains underexplored despite the unique risks of structured attributes and limited record diversity.
In this paper, we introduce \method, Membership Inference Attack via Error Prediction for Tabular Data, a novel black-box attack specifically designed to target tabular diffusion models. 
\method\ constructs error-based feature vectors by masking and reconstructing attributes of target records, disclosing membership signals based on how well these attributes are predicted. 
\method\ operates without access to the internal components of the generative model, relying only on its synthetic data output, and was shown to generalize across multiple state-of-the-art diffusion models.
We validate \method\ on three diffusion‑based synthesizers, achieving AUC‑ROC scores of up to 0.599 and TPR@10\% FPR values of 22.0\% in our internal tests. 
Under the MIDST 2025 competition conditions, \method\ achieved second place in the Black-box Multi-Table track (TPR@10\% FPR = 20.0\%).
% In an external evaluation under competition conditions, \method\ achieved second place.\footnote{We omit the name of the competition for anonymity during peer review.}
These results demonstrate that our method can uncover substantial membership leakage in synthetic tabular data, challenging the assumption that synthetic data is inherently privacy-preserving.
Our code is publicly available at \url{https://github.com/eyalgerman/MIA-EPT}.
\end{abstract}

%% file: content/introduction.tex
\begin{figure}[ht]
  \centering
  \includegraphics[width=\columnwidth]{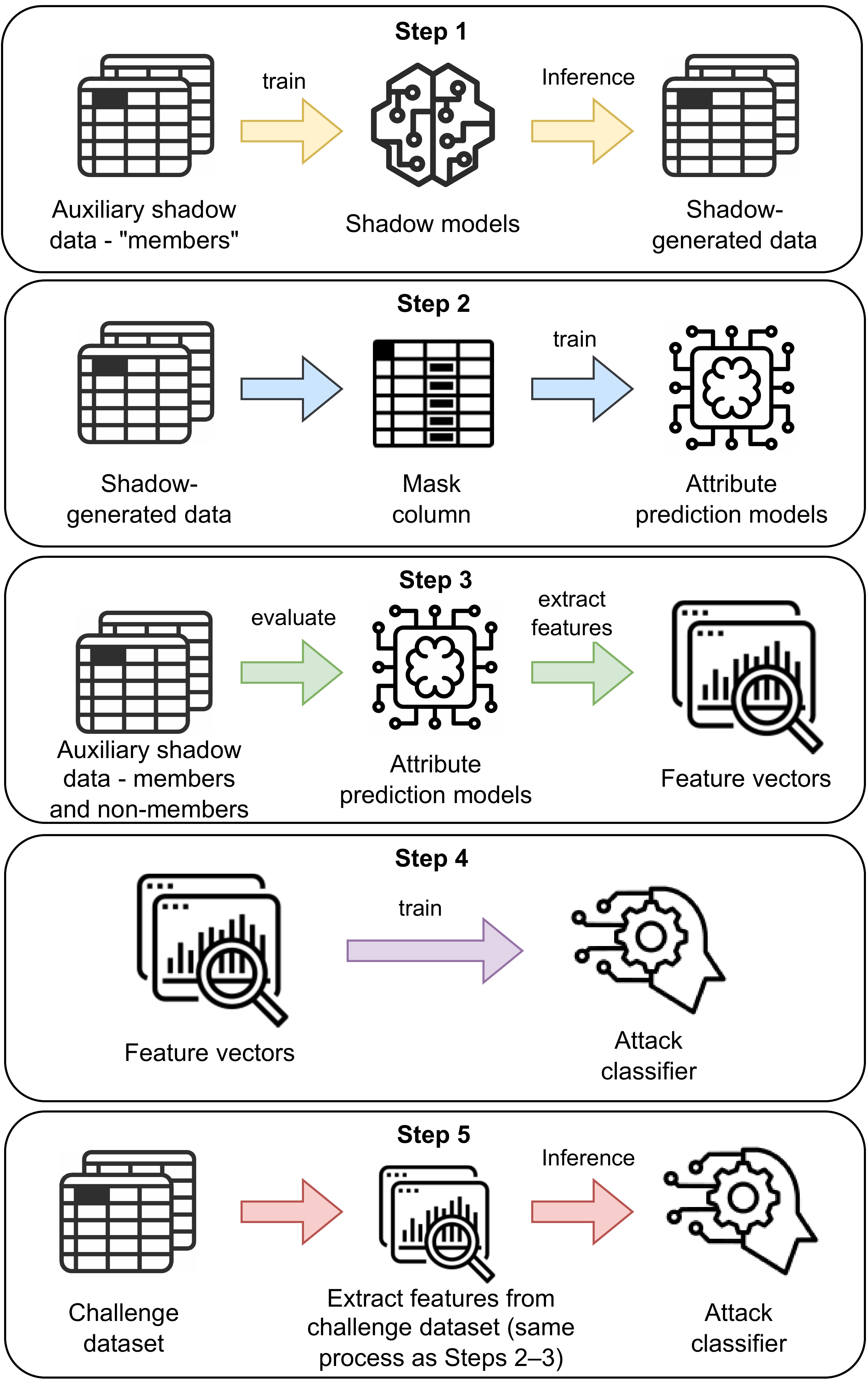}
  \caption{\method\ pipeline showing the flow through: (1) Shadow Model Training, (2) Attribute Prediction Model Training, (3) Feature Extraction, (4) Attack Classifier Training, and (5) Membership Prediction on the Challenge Dataset.}
  \label{fig:method}
\end{figure}

\section{Introduction}

Synthetic data generation has become an effective tool, capable of tackling challenges associated with data scarcity, privacy constraints, and data sharing restrictions. 
It enables model development and evaluation in scenarios where real-world data is limited, imbalanced, or inaccessible due to legal or ethical concerns~\cite{goncalves2020generation,talwar2020evaluating}. 
It also provides a practical means of simulating rare events, improving generalization, and performing rigorous validation~\cite{dash2020medical, pagano2023bias}. 

Synthetic data generation has thus become a promising solution, particularly in sensitive domains like healthcare and finance~\cite{villaizan2024diffusion}, due to its ability to enable data sharing while preserving privacy.
Diffusion models, a class of generative models originally designed for image synthesis, have demonstrated impressive capabilities in generating high-quality and diverse visual content~\cite{ho2020denoising,saharia2022photorealistic,rombach2022high}. 
These models have since been adapted to tabular data, yielding high-fidelity synthetic tables that mimic real datasets~\cite{kotelnikov2023tabddpm,zhang2023mixed}. 

However, even without access to the model's explicit training data, AI-generated synthetic data can inadvertently leak information about the real individuals whose data was used to train the generative model~\cite{fang2024understanding,shi2025comprehensivesurveysynthetictabular}. 
Membership inference attacks (MIAs) exploit such leakage by determining whether a given data sample was part of the training set of a model~\cite{Shokri2017}. 
This poses a serious privacy threat, especially in regulated domains (e.g., healthcare), where identifying whether an individual's data was used in training can reveal sensitive information such as their presence in a dataset or a medical condition, thereby violating synthetic data’s core promise of privacy.
The issue is especially pronounced for diffusion-based tabular synthesizers. 
These models capture complex data distributions and relationships (including mixed data types and relational tables), but their accuracy may rely on subtle traces of the original training data, potentially leading to privacy leakage~\cite{duan2023diffusion}. 

Existing MIA techniques may underperform on tabular diffusion models due to the heterogeneity of tabular data and the diffusion model’s unique sampling process. 
% The need for specialized attack strategies that address the privacy challenges of diffusion-generated tabular data motivated our development of \method\ (Membership Inference Attack via Error Prediction for Tabular Data). 
Evaluating the privacy risks of diffusion-generated tabular data requires specialized membership inference strategies tailored to their unique characteristics. 
In this paper, we present \method\ (Membership Inference Attack via Error Prediction for Tabular Data), a novel black-box MIA tailored for tabular diffusion models that relies solely on the synthetic data generated by these models.
% This motivated our development of \method\ (Membership Inference Attack via Error Prediction for Tabular Data).
% \method\ directly targets the privacy of diffusion-synthesized tables by leveraging a simple but powerful hypothesis: if a synthetic data generator has inadvertently memorized a training record, that record will be easier to predict (in terms of attribute values) than a record it never saw. 
\method\ is based on a fundamental hypothesis: if a synthetic data generator has inadvertently memorized a training sample, then that sample’s attribute values will be easier to predict from the synthetic data it generates than those of an unseen sample.
In other words, samples that were used to train the diffusion model will likely yield lower prediction errors on models trained on the synthetic data, compared to samples that were not used during training.
% tend to yield lower prediction errors under models trained on the synthetic data, compared to genuinely novel samples.
By exploiting this error-based signal, \method\ can effectively distinguish members from non-members in the generative model’s training set. 
\method\ contributes to a growing body of black-box MIA techniques targeting tabular diffusion models, distinguishing itself through its simplicity, broad applicability, and versatility across both single- and multi-table settings.
% generality, and support for both single- and multi-table settings. 

% Unlike prior work that often required auxiliary real-world data, shadow datasets, or diffusion-specific architectural assumptions, \method\ relies solely on the synthetic data generated. 
Unlike prior work that required direct access to target models, our method operates in a strict black-box setting—relying only on the synthetic data generated by the target model, while utilizing auxiliary shadow datasets and knowledge of the target architecture to simulate the generative process.
\method\ also handles heterogeneous tabular formats, accommodating both numerical and categorical columns.
By training per-column attribute prediction models to reconstruct masked attributes and extracting structured error profiles, our proposed method uncovers reliable signals of membership leakage.
Our approach follows the shadow-model paradigm and leverages reconstruction-based membership cues, adapting them to the unique challenges of tabular diffusion models and relational data.

% with MIDST
\method\ achieves strong empirical results in the MIDST 2025 challenge (Membership Inference over Diffusion-models-based Synthetic Tabular data) at SaTML 2025~\cite{MIDST2025}. 
In this rigorous benchmark, \method\ ranked second in the Black-box Multi-Table track, achieving a true positive rate of up to 20\% at a false positive rate of 10\%. 
These results demonstrate that \method\ reveals measurable privacy leakage in state-of-the-art (SOTA) tabular diffusion models and highlight the need for robust privacy defenses in synthetic data generation.

% after remove MIDST name
% \method\ achieves strong empirical results in a public benchmark competition on membership inference on diffusion-generated tabular data, held as part of a leading academic conference.\footnote{Details of the benchmark and competition will be disclosed in the camera-ready version to preserve anonymity.} \method\ ranked second in the competition, achieving a true positive rate of up to 20\% at a false positive rate of 10\%.
% These results demonstrate that \method\ reveals measurable privacy leakage in state-of-the-art (SOTA) tabular diffusion models and highlight the need for robust privacy defenses in synthetic data generation.

% In this paper, we describe \method's design, its theoretical underpinnings, and the empirical evaluation performed on a public benchmark for membership inference on diffusion-generated tabular data.\footnote{Details of the benchmark will be disclosed in the camera-ready version to preserve anonymity.}
% On this rigorous benchmark, \method\ ranked second, achieving a true positive rate of up to 20\% at a false positive rate of 10\%. 
% The results, which demonstrate that \method\ reveals measurable privacy leakage in state-of-the-art (SOTA) tabular diffusion models, underscore the importance of robust privacy defenses for synthetic data.

% Our work contributes to the growing evidence that synthetic data is not a guaranteed privacy panacea. 
Diffusion models, despite their complexity and impressive ability to generate realistic tabular data, are susceptible to membership leakage as exposed by attacks like \method. 
The immediate implication is that organizations using diffusion-based data synthesis (or any generative model) for privacy should rigorously evaluate their outputs for such leaks.
\method\ reveals how subtle error-based signals embedded in synthetic tabular data can be leveraged to infer membership in the original training data.
% It underscores the need for a privacy–utility balance: maximizing the utility of synthetic data (by retaining important patterns) inevitably runs the risk of retaining traces of individual data points. 
It also highlights the tradeoff between privacy and utility, as maximizing the utility of synthetic data by preserving important patterns inevitably increases the risk of retaining traces of individual data points.

In summary, this paper makes the following contributions:
\begin{itemize}
\item We introduce \method, a novel black-box MIA tailored for tabular diffusion models, which operates solely on the synthetic heterogeneous tabular data they generate. % that relies on just the synthetic heterogeneous tabular data generated. % and accommodates heterogeneous tabular structures with both numerical and categorical attributes.

\item We propose a generalizable pipeline that uses per-column attribute prediction models to reconstruct masked attributes and extract structured error profiles, effectively exposing membership leakage across single-table and multi-table settings without relying on internal model access.
% \item We demonstrate \method’s practical impact in MIDST 2025—ranking 2\textsuperscript{nd} in the Black‑box Multi‑Table track and uncovering tangible privacy risks in modern synthetic‑data generators.
% \item We demonstrate \method’s practical effectiveness in a public competition on membership inference over synthetic tabular data—achieving second place.
% \item We demonstrate \method’s practical effectiveness through both internal evaluations and a public competition on membership inference on synthetic tabular data—achieving AUC-ROC scores of up to 0.599 and TPR@10\% FPR of 22.0\% internally, and securing second place in the competition.
\item We demonstrate \method’s practical effectiveness through both internal evaluations and the MIDST 2025 challenge on membership inference over diffusion-based synthetic tabular data—achieving AUC-ROC scores of up to 0.599 and TPR@10\% FPR of 22.0\% internally, and securing second place in the Black-box Multi-Table track of the competition.
% \footnote{Details of the competition and ranking will be disclosed in the camera-ready version to preserve anonymity.}

\end{itemize}

%% file: content/related_work.tex
\section{Related Work}
%As diffusion methods for synthetic tabular data mature, there is a need to understand their privacy vulnerabilities, particularly to MIAs.

\subsection{Diffusion Models for Tabular Data Generation}

Diffusion models~\cite{ho2020denoising} have recently emerged as a powerful class of generative models; originally developed for image synthesis, these models are increasingly being adapted for structured data domains such as tabular data. 
Compared to generative adversarial networks (GANs)~\cite{xu2019modeling,choi2017medgan} and variational autoencoders (VAEs)~\cite{xu2019synthesizing}, diffusion models offer superior sample quality and training stability.
These properties reduce the risk of overfitting and unintended memorization, making diffusion models an increasingly attractive choice for generating synthetic data in privacy-sensitive domains~\cite{shi2025comprehensivesurveysynthetictabular}.

Novel methods have been proposed in studies aimed at extending diffusion models to the tabular domain and addressing the unique challenges posed by structured data.
TabDDPM~\cite{kotelnikov2023tabddpm} is a denoising diffusion probabilistic model specifically designed for tabular data, incorporating architectural and loss modifications to handle heterogeneous feature types, including continuous, categorical, and ordinal attributes. 
Building on this direction, TabSyn~\cite{zhang2023mixed} introduced architectural enhancements and conditional generation mechanisms to support controllable sampling and improve the modeling of complex tabular distributions. 
More recently, ClavaDDPM~\cite{pang2024clavaddpm} advanced the field by leveraging clustering labels as intermediate representations to capture inter-table relationships. 
This approach explicitly enforces foreign key constraints, thereby facilitating the generation of high-quality multi-table tabular data.
% CtrTab~\cite{li2025ctrtab} 
As tabular diffusion models become increasingly sophisticated, understanding their privacy implications,including potential memorization, has become an important research direction~\cite{fang2024understanding}.

\input{content/method_algorithm}

%  new
\subsection{Membership Inference Attacks}
MIAs exploit statistical differences between a model’s behavior on training versus unseen data. 
Shokri \etal~\cite{Shokri2017} first demonstrated classifier‐based MIAs using shadow models, showing that overfitted classifiers output higher confidence scores for training inputs. 
Overfitted classifiers also typically output higher softmax probabilities for training inputs than for unseen examples, enabling attackers to infer membership based on output entropy or confidence~\cite{Shokri2017,yeom2018privacy,hu2022membership}. 
As research in this area advanced, the evolving techniques required less information: for example, label-only attacks~\cite{sablayrolles2019white} rely solely on the model’s predicted label yet are capable of detecting statistical differences between training and hold-out data. 

%LLMs
In the context of large language models (LLMs), subsequent work quantified memorization risks, highlighting trade-offs between utility and privacy~\cite{carlini2022quantifying}.
Tag\&Tab, a keyword-based MIA, was proposed to detect whether specific pretraining data was used in LLMs~\cite{antebi2025tag} by analyzing the model’s output for signs of memorized keywords indicative of pretraining data exposure.
% \citet{carlini2022quantifying} examined the extent of memorization in neural language models, highlighting the nuanced trade-off between utility and privacy risks. 
% More recently,~\citet{antebi2025tag} introduced Tag\&Tab, a keyword-based MIA specifically designed to detect whether certain pretraining data was used in large language models.

%Images
MIAs have also been extended to image generative models.
The LOGAN attack leverages the discriminator of a GAN to distinguish between real and synthetic samples~\cite{hayes2017logan}, while other methods have used Monte Carlo sampling and reconstruction-error techniques to attack VAEs~\cite{hilprecht2019monte}.
A membership error metric was later introduced to refine these attacks and improve detection accuracy~\cite{shafran2021membership}.
% Subsequent work extended MIAs to generative models for images. The LOGAN attack proposed by~\citet{hayes2017logan} leverages a GAN’s discriminator to distinguish real from synthetic samples, while~\citet{hilprecht2019monte} used Monte Carlo and reconstruction‐error methods against VAEs. 
% \citet{shafran2021membership} refined this by defining a “membership error” metric, improving detection accuracy. 

%multimodal
In the multi-modal setting, recent work has explored MIAs targeting large-scale models~\cite{li2024identity, hu2022m}.
Some methods have focused on realistic attacker constraints and transferability across architectures~\cite{ko2023practical}, while others have shown that variance in generated captions can be used to detect membership without requiring model internals~\cite{samira2025variance}.
% Recently, researchers have proposed approaches for MIAs targeting multimodal models~\cite{li2024identity, hu2022m}.
% \citet{ko2023practical} conducted a pilot study on practical MIAs targeting large-scale multimodal models; the researchers focused on realistic attacker constraints and transferability across architectures. 
% \citet{samira2025variance} proposed a variance-based attack against image captioning systems, showing that generative variance can be a strong signal for membership, even in the absence of model internals.

%Diffusion models
With the rise of diffusion models~\cite{ho2020denoising}, MIAs have faced new challenges: these models generate data through iterative denoising processes and lack explicit discriminators, unlike GANs.
SecMI traces reconstruction errors across sampling steps~\cite{duan2023diffusion}, and other methods have proposed extraction attacks directly on diffusion checkpoints~\cite{carlini2023extracting}.
% \citet{duan2023diffusion} introduced SecMI, which traces reconstruction errors across sampling steps, and~\citet{carlini2023extracting} proposed extraction attacks on diffusion checkpoints.

In the tabular domain, density-based attacks such as DOMIAS~\cite{vanbreugel2023membershipinferenceattackssynthetic} detect overfitting by identifying regions with unusually high probability mass. 
Fully synthetic attack pipelines have been developed, where a generator is queried to produce shadow datasets, followed by retraining a new generator on the resulting synthetic data. Membership is then inferred by comparing how well the target record is reconstructed when held out~\cite{guepin2023synthetic}.
% \citet{guepin2023synthetic} advanced this idea by proposing a fully synthetic attack pipeline in which the generator is queried to create shadow datasets, a new generator is retrained on the released synthetic outputs, and the target record is held out when constructing training samples. 
Building on the insight that auxiliary data may not be necessary,Wu \etal~\cite{wu2025winningmidstchallengenew} introduced a MLP that learns noise- and timestep-aware features directly from diffusion losses.
To the best of our knowledge, they were the first to conduct MIAs against tabular diffusion models, showing that synthetic-only attacks can rival traditional approaches requiring additional real data.

However, none of these approaches exploit the structural dependencies and mixed-type relationships present in multi-table data. 
In contrast, our proposed attack incorporates attribute-wise reconstruction error patterns, enabling more precise membership inference in complex synthetic tabular datasets.
Unlike the approach proposed by Wu \etal, which originated as a white-box attack and was later adapted to a black-box setting by retraining new generators on synthetic data, our method operates directly on the synthetic data generated without the need to train additional diffusion models for each target. 
This makes our attack entirely model-agnostic—it does not require access to the original generative process or knowledge of the specific model architecture, relying solely on the available synthetic tables.

%% file: content/method_algorithm.tex
% ----------------------------
% Algorithm Block 
% ----------------------------
% \begin{algorithm}[!htbp]
% \begin{algorithm}[ht]
\begin{algorithm}[!tb]
\caption{MIA-EPT Pipeline}
\label{alg:mia-ept}
% \KwIn{
%   Synthetic data from shadow and target models: $\mathcal{S}_{\text{shadow}}, \mathcal{S}_{\text{target}}$;\\
%   Shadow model datasets: member data $\mathcal{D}_{\text{member}}$, non-member data $\mathcal{D}_{\text{non-member}}$;\\
%   Challenge dataset: $\mathcal{C}$
% }
\KwIn{
  Auxiliary shadow data $\mathcal{D}_{\text{aux}}$;\\
  Synthetic data from the target model: $\mathcal{S}_{\text{target}}$;\\
  Challenge dataset: $\mathcal{C}$
}
\KwOut{Membership scores for each record in $\mathcal{C}$}
\textbf{Step 1: Train Shadow Diffusion Models and Generate Synthetic Data}\\
Partition $\mathcal{D}_{\text{aux}}$ into $2M$ disjoint splits\\
\ForEach{shadow model $i = 1, \ldots, M$}{
    Select $\mathcal{D}_{\text{member}}^{(i)}$ as one split (used as member data for $G_i$)\\
    Set $\mathcal{D}_{\text{non-member}}^{(i)}$ as another split (used as non-member data for $G_i$)\\
    Train shadow diffusion model $G_i$ on $\mathcal{D}_{\text{member}}^{(i)}$\\
    Generate synthetic dataset $\mathcal{S}_i = G_i(\mathcal{N})$\\
}

\textbf{Step 2: Train Attribute Prediction Models on Synthetic Data} \\
\ForEach{dataset $\mathcal{S}_i$ in $\mathcal{S}_{\text{shadow}} $}{
    \ForEach{column $c$ in $\mathcal{S}_i$}{
        Train attribute prediction model $H_c^{\mathcal{S}_i}$ to predict column $c$ from $\mathcal{S}_i \setminus c$
    }
}

\textbf{Step 3: Extract Features from Shadow Models using Attribute Prediction Models} \\
\ForEach{dataset $\mathcal{D}_i$ in \{$\mathcal{D}_{\text{member}} \cup \mathcal{D}_{\text{non-member}} \} $}{
    \ForEach{$x_j \in \mathcal{D}_i$, each column $c$}{
        Predict $\hat{x}_{j,c} = H_c^{\mathcal{S}_i}(x_{j} \setminus c)$ \\
        Extract and aggregate prediction errors into feature vector $f_{i,j}$
    }
}

\textbf{Step 4: Train Attack Classifier} \\
\ForEach{dataset $\mathcal{D}_i$ in $\{\mathcal{D}_{\text{member}} \cup \mathcal{D}_{\text{non-member}} \}$}{
    \ForEach{feature vector $f_{i,j}$ extracted from $\mathcal{D}_i$}{
        Assign label $y_{i,j} \gets 
        \begin{cases}
            1 & \text{if } \mathcal{D}_i \in \mathcal{D}_{\text{member}}\\
            0 & \text{otherwise}
        \end{cases}$
    }
}
Train attack model $\mathcal{A}$ on the feature-label pairs $\{(f_{i,j}, y_{i,j})\}$

\textbf{Step 5: Evaluate Membership on the Challenge Dataset} \\

\ForEach{column $c$ in $\mathcal{S}_{\text{target}}$}{
    Train attribute prediction model $H_c^{\mathcal{S}_{\text{target}}}$ to predict column $c$ from $\mathcal{S}_{\text{target}} \setminus c$
}

\ForEach{$x_j \in \mathcal{C}$}{
    \ForEach{column $c$}{
        Predict $\hat{x}_{j,c} = H_c^{\mathcal{S}_{\text{target}}}(x_j \setminus c)$ and compute prediction error
    }
    Aggregate prediction errors into feature vector $f_j$\\
    Compute membership score $\mathcal{A}(f_j)$
}

\end{algorithm}

%% file: content/method.tex
\section{\label{sec:method}Method}
We introduce \method\, a novel MIA targeting synthetic data generation models trained on tabular data.
\method\ follows the shadow-model paradigm introduced by Shokri \etal~\cite{Shokri2017} and leverages reconstruction-based membership cues, inspired by recent work on generative models~\cite{carlini2023extracting}. 

Our attack operates in a black-box setting, as it does not query or interact with the target generative model, relying solely on the synthetic data generated~\cite{guepin2023synthetic}.
Our approach can be adapted to various types of tabular generative models, wherever the hypothesis that members are predicted with lower error holds.
% Our approach can be adapted for different types of tabular generative models – anywhere the hypothesis "members are predicted with lower error" holds~\cite{hilprecht2019monte}.
An overview of our workflow is provided in Figure~\ref{fig:method}, with its detailed steps presented in Algorithm~\ref{alg:mia-ept}.

\subsection{Attacker Model}
\method\ operates in a pure black-box setting, where the adversary only has access to the synthetic dataset released by the target tabular generator and is unable to make queries to the model itself. 
The attacker is assumed to possess an auxiliary shadow dataset with similar properties (i.e., drawn from the same distribution) to the target model’s training data; this enables the adversary to train shadow diffusion models which are used to simulate the victim’s generative process. 
Additionally, we assume that the attacker knows the target model’s architecture, enabling them to train their own synthetic data generators using the same architecture.
The attacker's goal is to determine the membership status of records in a challenge dataset $\mathcal{C}$—a set of candidate records whose inclusion in the target model's training data is unknown.

Our pipeline is comprised of five main steps: (1) Shadow Model Training, (2) Attribute Prediction Model Training, (3) Feature Extraction (Error Profiles), (4) Attack Classifier Training, and (5) Membership Prediction on the Challenge Dataset.

\subsection{Shadow Model Training}
The auxiliary shadow data $\mathcal{D}_{\mathrm{aux}}$ is used to train shadow diffusion models, each simulating the target model’s generative process. 
In this step, for each shadow model, we also construct a labeled dataset comprising “member” records (used to train the model) and “non-member” records (excluded from the training set). 
Synthetic datasets generated by these shadow models serve as the basis for subsequent attack steps.

\subsection{Attribute Prediction Model Training}
The \emph{attribute prediction models} are trained on synthetic datasets generated by the shadow models.
Each attribute prediction model is trained to predict the value of a specific attribute (column) given the values in the remaining columns, using either regression or classification (depending on the attribute type). 
Given the synthetic dataset $\mathcal{S}_i$ generated by shadow model $i$, we train a set of models $\{H_c^{\mathcal{S}_i}\}$, each tasked with predicting column $c$ from $\mathcal{S}_i \setminus c$ (treating $c$ as masked). 
These models capture inter-column relationships as encoded by the shadow models. 
By training only on synthetic data, the attribute prediction models learn the underlying data distribution as represented by the generative (shadow) model. 

\subsection{Feature Extraction (Error Profiles)}
We extract error-based features using the trained attribute prediction models. 
In this step, we apply each attribute prediction model to records used in the shadow model setup, where the membership status is known to the adversary; the attribute prediction models are applied to both member records and non-member records.

For each record $x_j$ in $\mathcal{D}_{\text{member}}$ or $\mathcal{D}_{\text{non-member}}$ associated with a given shadow model, and for each attribute prediction model $H_c^{\mathcal{S}_i}$, we mask column $c$ and use the model to predict the corresponding cell value. 
The prediction error or correctness for each column is aggregated into a feature vector $f_j$ for record $x_j$, collectively forming the record’s error profile across all attributes.
Specifically, we extract the following five types of error-based features:

\begin{itemize}
    \item \textbf{Actual value}: The true value of the column in the record $x_{j,c}$. 
    This provides context that may help the attack classifier learn whether certain values are memorized more accurately.
    
    \item \textbf{Predicted value}: The output produced by the attribute prediction model $\hat{x}_{j,c}=H_c^{S_i}(x_{j}\setminus c)$, which attempts to reconstruct the masked column. 
    This feature applies to both continuous and categorical columns (as either a numeric prediction or a predicted category).
  
    \item \textbf{Error}: For continuous (regression) columns, we compute the absolute prediction error \\$| x_{j,c}-H_c^{S_i}(x_{j}\setminus c)|$, which captures how far the predicted value is from the true value.
    
    \item \textbf{Error ratio}: For continuous (regression) columns, we normalize the absolute error by the magnitude of the actual value itself.
    This results in the relative error, which reflects how large the prediction error is compared to the true value, although care must be taken with values near zero.
    
    \item \textbf{Accuracy}: For categorical (classification) columns, we assign a binary indicator, which is '1' if the predicted category matches the true category, and '0' otherwise. This feature indicates whether the prediction was correct.
\end{itemize}

The core intuition is that if $x_j$ was included in the training data of the (shadow) diffusion model, the corresponding attribute prediction models trained on synthetic data will reconstruct the attributes of $x_j$ more accurately. 
In contrast, non-member records will, on average, result in higher reconstruction errors or misclassifications, thus providing a reliable indication of membership inference.

\subsection{Attack Classifier Training}
Using the extracted features, we train an attack classifier $\mathcal{A}$ to distinguish members from non-members. 
Each feature vector $f_{i,j}$ is labeled according to the membership status of its originating dataset:
$y_{i,j} = 1$ for $\mathcal{D}_i \in \mathcal{D}_{\text{member}} $, else $y_{i,j} = 0$.
We experiment with various classifiers, selecting the best-performing one on a validation set. 
The attack model ultimately learns to map error profiles to membership likelihoods, outputting a membership score that indicates the probability that a given record was included in the target model's training data.

\subsection{Membership Prediction on the Challenge Dataset}
Finally, we run the trained attack classifier on each record in the challenge dataset $\mathcal{C}$ to determine membership status.
Specifically, we train attribute prediction models $H_c^{\mathcal{S}_{\text{target}}}$ on $\mathcal{S}_{\text{target}}$, and for each record $x_j \in \mathcal{C}$ we predict the value for each column $c$ to compute the associated prediction errors. 
These are aggregated into a feature vector $f_j$, which is fed to the attack classifier to obtain a membership score $\mathcal{A}(f_j)$. 
A high score suggests that $x_j$ is likely a member of the training set. 
These scores can be thresholded to produce binary decisions or used to rank the challenge records by membership likelihood.

%% file: content/evaluation.tex
% \section{Evaluation}
\section{Evaluation}
For full hardware and software configurations, including GPU/CPU specifications, and library versions, see Appendix: \textit{Training and Evaluation Configurations}.
% Appendix~\ref{appendix:training}.

\subsubsection*{Target Models.} 
We evaluate our method on three SOTA tabular diffusion models:TabDDPM~\cite{kotelnikov2023tabddpm}, TabSyn~\cite{zhang2023mixed}, and ClavaDDPM~\cite{pang2024clavaddpm}, using synthetic datasets produced by each.
For each diffusion model (TabDDPM, TabSyn, ClavaDDPM), we generate synthetic data from $30$ model instances. 
We use the first $25$ instances to train our attribute prediction models, extract features, and train the attack classifier, while reserving the remaining $5$ for evaluation. 
Each instance yields a synthetic table with the same schema; we label records from the synthetic data model’s own training set as \emph{members} and sample an equal-sized hold-out set of \emph{non-members} from unseen real data.
% Importantly, our attack operates in a pure black-box setting: it does not require access to the internals or the checkpoints of the generative models themselves, relying solely on the released synthetic data.
All datasets and pretrained checkpoints used in our experiments were provided by the organizers of the MIDST 2025 challenge~\cite{MIDST2025}.
% All datasets and pretrained checkpoints used in our experiments were obtained from a public benchmark.\footnote{Details on the benchmark and its source will be provided in the camera-ready version to preserve anonymity.}

\subsubsection*{Metrics.} 
We evaluate attack performance using two standard metrics: 
(i) AUC-ROC, the area under the ROC curve (0.5 = random, 1 = perfect), which gauges threshold-independent separability; and  
(ii) TPR@10\% FPR, the true positive rate achieved when the false positive rate is capped at 10\%, representing a realistic operating point, as used by~\cite{carlini2022membership}.
% All runs use a fixed random seed of 42.

\subsubsection*{Attack Classifiers.}  
In our experiments, we evaluate two classifiers: XGBoost and CatBoost. Each classifier is trained separately on the extracted error profiles, and only one is used per run. We compare their performance to identify the most effective choice for membership inference on each model.
% We compare two classifiers on the extracted error profiles: XGBoost and CatBoost.
All models are trained on the concatenated profiles from the $25$ training instances and evaluated on the $5$ test instances.

\subsubsection*{Feature Extraction.}  
We train one attribute prediction model per column (classification for categorical, regression for continuous) in each synthetic table and extract an error profile for each record. 
We experiment with different subsets of feature types (actual values, predicted values, errors, error ratios, and categorical-accuracy indicators) by concatenating selected features into the feature vectors.
% In our main results, we report on the full error profile, which includes all five feature categories.

%% file: content/results.tex
\section{Results}

% \begin{figure*}[ht]
%     \centering
%     \includegraphics[width=0.95\textwidth]{content/figures/roc_auc_with_zoom_models_comparison_20250716_101137.png}
%     \caption{
%         ROC-AUC curves for TabDDPM, TabSyn, and ClavaDDPM on the test set.
%         The main plot shows the full ROC curve for each model, while the inset zooms in on the critical region with a false positive rate (FPR) and true positive rate (TPR) below 0.2, enabling fine-grained comparison at low error rates. The dashed line denotes random guessing.
%     }
%     \label{fig:roc_auc_with_zoom}
% \end{figure*}

\begin{figure*}[t]
    \centering
    \includegraphics[width=0.95\linewidth]{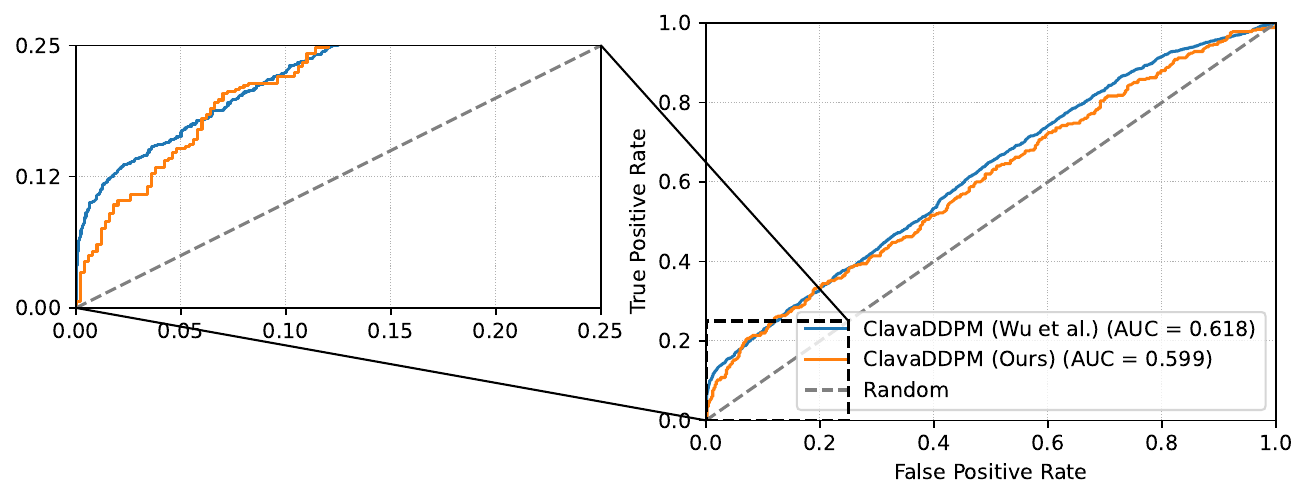}
    % \caption{
    % ROC curves for ClavaDDPM, comparing our method to the baseline of Wu et al. The inset zooms in on the low-FPR region.
    % }
    \caption{
    The ROC curves for the multi-table model ClavaDDPM on the test set, comparing \method\ with the recent baseline of Wu \etal~\cite{wu2025winningmidstchallengenew}. The inset zooms in on the low-FPR region.
    }
    \label{fig:roc_auc_clavaddpm}
\end{figure*}

\begin{figure*}[t]
    \centering
    \includegraphics[width=0.95\linewidth]{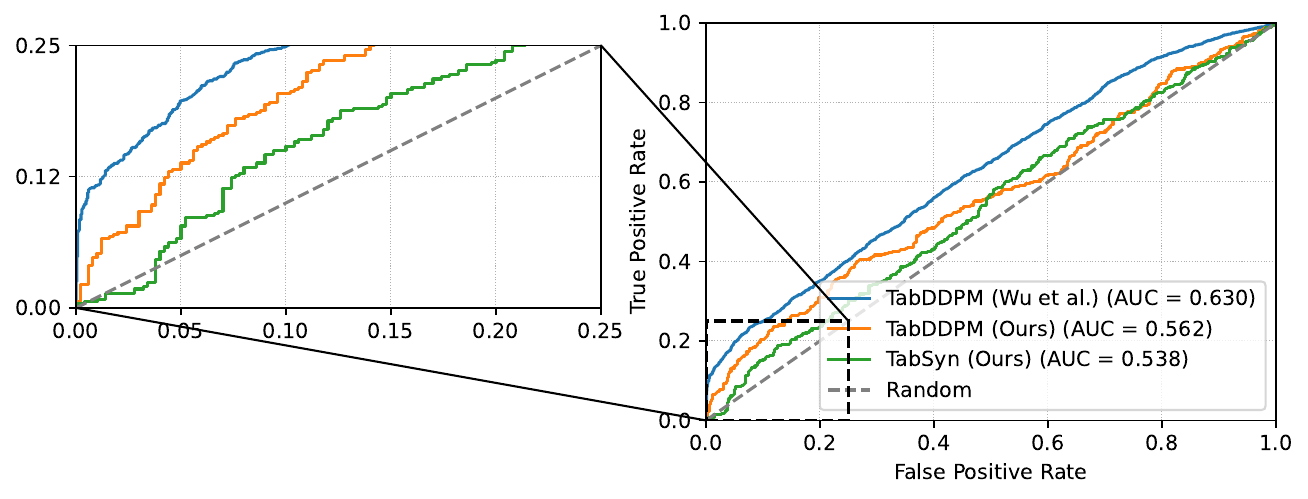}
    % \caption{
    % ROC curves for TabDDPM and TabSyn models, comparing our method and to the baseline of Wu et al. The inset zooms in on the low-FPR region. }
    \caption{
    The ROC curves for the single-table models TabDDPM and TabSyn on the test set, comparing \method\ with the recent baseline of Wu \etal~\cite{wu2025winningmidstchallengenew}. The inset zooms in on the low-FPR region.
    }
    \label{fig:roc_auc_tabddpm}
\end{figure*}

\input{tables/top-5_results}

Figures~\ref{fig:roc_auc_clavaddpm} and~\ref{fig:roc_auc_tabddpm} present the ROC curves for TabDDPM, TabSyn, and ClavaDDPM on the test set, comparing the performance of our attack with the baseline of Wu \etal~\cite{wu2025winningmidstchallengenew}. 
For each model, we report and plot the best-performing configuration in terms of  the TPR@10\% FPR metric.
As seen in Figure~\ref{fig:roc_auc_tabddpm}, TabDDPM (Wu et al.) achieves a TPR@10\% FPR of 0.249 and an AUC of 0.630, while our attack achieves a TPR@10\% FPR of 0.204 and an AUC of 0.562. 
Since Wu et al. did not evaluate or implement their attack on the TabSyn model, only our attack is evaluated on this model, achieving a TPR@10\% FPR of 0.150 and an AUC of 0.538, 
As seen in Figure~\ref{fig:roc_auc_clavaddpm}, ClavaDDPM (Wu et al.) achieves a TPR@10\% FPR of 0.224 and an AUC of 0.618, while our attack achieves a TPR@10\% FPR of 0.22 and an AUC of 0.599. 

The insets in both figures zoom in on the low-FPR region, which is critical for real-world MIAs. 
These plots highlight that both attacks are able to detect members with a high true positive rate, even with low false positive rates, underscoring the privacy risk in synthetic tabular data. 
On the ClavaDDPM model, our attack outperforms the attack of Wu et al., achieving a higher TPR with very low FPR in some cases, and overall, our attack achieves performance comparable to Wu et al., as seen in Figure~\ref{fig:roc_auc_clavaddpm}. 
Achieving a high TPR at very low FPR is particularly important in practical privacy scenarios, where defenders often require MIAs to operate effectively at low FPR thresholds in order to avoid a large number of false alarms.

% With MIDST
% These results confirm our earlier table-based findings and show that our error-profile approach can capture subtle memorization signals embedded in synthetic tables even under strict black-box constraints, with strong TPR@FPR values. 
% Under the MIDST 2025 challenge rules~\cite{MIDST2025}, which enforce additional hidden test tables and strict time constraints, \method\ secured 2nd place in the Black‑box Multi‑table track. 
% Our official challenge scores—computed on the organizer’s blind test set—were slightly lower but still competitive, with a TPR@FPR=10\% of 20.0\%.

% without MIDST
These results show that our attack can capture subtle memorization signals embedded in synthetic tables even under strict black-box constraints, with strong TPR@FPR values.
% \method\ came in second place on a public benchmark organized as part of a conference competition.
Under the MIDST 2025 challenge rules~\cite{MIDST2025}, which enforce additional hidden test tables and strict time constraints, \method\ secured 2nd place in the Black‑box Multi‑table track. 
Our official competition scores, which were computed on a blind test set provided by the organizers, resulted in a TPR@10\% FPR of 20.0\%.

While the method of Wu \etal achieves even stronger membership inference results on the same benchmark, their approach begins as a white-box attack and transitions to a black-box scenario by first training new diffusion models on the synthetic data generated by the target model, essentially building a custom generative pipeline for each target case.
In contrast, our method achieves competitive membership inference simply by analyzing reconstruction errors from attribute
prediction models trained on the generated synthetic data, without the need to retrain or access a new diffusion model for each target model. 
This makes our approach much more lightweight, easier to deploy, and truly model-agnostic.
% , and applicable in settings where retraining generative models is impractical or infeasible, assuming that shadow models have been trained in advance.
While the results of Wu \etal underscore the vulnerability of diffusion models to more specialized attacks, our findings show that even straightforward error-based probing can uncover substantial privacy leakage, highlighting the severity of membership risks associated with synthetic tabular data.

\subsection{Feature Ablation Analysis}
To evaluate the contribution of different components in the error profile, we systematically analyze five feature types: \emph{actual value}, \emph{predicted value}, \emph{error}, \emph{error ratio}, and \emph{accuracy}.
Table~\ref{tab:top5_models} lists the top-five feature–classifier configurations for each diffusion model, ranked by the TPR@10\% FPR value.

On \textbf{TabDDPM}, the highest TPR@10\% FPR (0.204) is achieved using CatBoost, with a lean feature set consisting of: \emph{actual}, \emph{error}, and \emph{accuracy}. 
The best AUC-ROC (0.567) is also obtained using CatBoost but with a larger feature set consisting of: \emph{actual}, \emph{error}, \emph{error ratio}, \emph{accuracy}, and \emph{prediction}. 
% Across all top-performing configurations, removing error-based features (\emph{error}, \emph{error ratio}) generally led to a reduction in both TPR and AUC, confirming their importance for continuous attribute membership inference.

On \textbf{TabSyn}, CatBoost with a feature set consisting of \emph{actual}, \emph{error ratio}, and \emph{prediction} yields the best TPR@10\% FPR (0.150), while XGBoost with a feature set of \emph{actual}, \emph{error}, and \emph{prediction} produces the top AUC-ROC (0.543). 
Interestingly, several compact feature sets performed competitively, suggesting that on the TabSyn generator, high membership signals can be captured, even with fewer features, particularly those emphasizing prediction and relative error.

On \textbf{ClavaDDPM}, the significance of the accuracy feature is evident: CatBoost with just the  \emph{actual} and \emph{accuracy} features achieves both the highest TPR@10\% FPR (0.220) and the highest AUC-ROC (0.599). 
Other top configurations also consistently include the accuracy feature, highlighting its key importance for this model’s schema, which is dominated by categorical features.

% For ClavaDDPM, the most significant takeaway is the dominance of the \emph{accuracy} for categorical data. The best AUC-ROC (0.599) resulted from CatBoost using only \emph{error} and \emph{accuracy}, while the highest TPR@10\% FPR (0.220) was achieved by a simple combination of \emph{actual value} and \emph{accuracy}. This suggests that ClavaDDPM's categorical-heavy schema amplifies the value of accuracy-based features compared to error-based ones.

Overall, these findings demonstrate that while absolute and relative error measures remain central for numeric attributes, categorical accuracy is a decisive feature in models like ClavaDDPM. 
Thus, an adaptive feature design that leverages schema-specific characteristics is crucial for robust membership inference.

%% file: tables/top-5_results.tex
\begin{table*}[!htbp]
\centering
% \small
\begin{tabularx}{\textwidth}{l l X c c}
\toprule
\textbf{Model} & \textbf{Classifier} & \textbf{Feature set} & \textbf{TPR@10\% FPR} & \textbf{AUC-ROC} \\
\midrule
\multirow{5}{*}{TabDDPM} & CatBoost & actual + error + accuracy & \textbf{0.204} & 0.562 \\
 & CatBoost & actual + error & 0.192 & 0.551 \\
 & XGBoost & actual + error + error\_ratio + accuracy + prediction & 0.190 & 0.550 \\
 & CatBoost & actual + error\_ratio + accuracy & 0.188 & 0.548 \\
 & CatBoost & actual + error + error\_ratio + accuracy + prediction & 0.184 & \textbf{0.567} \\
\midrule
\multirow{5}{*}{TabSyn} & CatBoost & actual + error\_ratio + prediction & \textbf{0.150} & 0.538 \\
 & CatBoost & actual + prediction & 0.146 & 0.530 \\
 & XGBoost & actual + prediction & 0.138 & 0.538 \\
 & CatBoost & actual + error\_ratio + accuracy + prediction & 0.138 & 0.533 \\
 & XGBoost & actual + error + prediction & 0.136 & \textbf{0.543} \\
\midrule
\multirow{5}{*}{ClavaDDPM} & CatBoost & actual + accuracy & \textbf{0.220} & \textbf{0.599} \\
 & CatBoost & actual + error\_ratio + accuracy & 0.200 & 0.574 \\
 & CatBoost & actual + accuracy + prediction & 0.198 & 0.599 \\
 & CatBoost & actual + error + error\_ratio + accuracy & 0.198 & 0.574 \\
 & CatBoost & actual + error + accuracy & 0.194 & 0.590 \\
\bottomrule
\end{tabularx}
\caption{Top-five feature configurations for each diffusion model, ranked by the TPR@10\% FPR value. The best TPR@10\% FPR and AUC-ROC for each model appear in bold.}
\label{tab:top5_models}
\end{table*}

%% file: content/conclusion.tex
\section{Conclusion}
We introduced \method, a black‑box MIA that targets diffusion-based synthetic tabular data. By training per-column predictors on synthetic tables, \method\ constructs error profiles that enable effective member/non-member classification.
% with MIDST
% Our empirical evaluation on the MIDST 2025 benchmark demonstrates that even without any access to model internals or auxiliary real data, \method\ achieves non‑trivial separation between members and non‑members across three SOTA tabular diffusion models (TabDDPM, TabSyn, and ClavaDDPM), with AUC‑ROC up to 0.599 and TPR@FPR=10\% of 22.0\%. 
% These results, while slightly below the most specialized synthetic‑only attacks in the literature, underscore that simple error‑profile probing is sufficient to expose substantial privacy leakage in modern synthetic‑data generators.
% Our method also proved highly effective in the MIDST 2025 competition, securing 2nd place in the Black‑box Multi‑Table track.
% Without MIDST
Our evaluation on a public benchmark demonstrates that even without any access to model internals or auxiliary real data, \method\ achieves non‑trivial separation between members and non‑members on three SOTA tabular diffusion models (TabDDPM, TabSyn, and ClavaDDPM), achieving an AUC‑ROC of up to 0.599 and a TPR@FPR=10\% of 22.0\%. 
These results highlight the privacy risks posed by synthetic data, even in black-box settings.
% These results, while slightly below the most specialized synthetic‑only attacks in the literature, underscore that simple error‑profile probing is sufficient to expose substantial privacy leakage in modern synthetic‑data generators.
% \method\ also proved highly effective in a conference-hosted competition on synthetic tabular data,
% \footnote{Details of the competition and ranking will be disclosed in the camera-ready version to preserve anonymity.}
% where it secured 2\textsuperscript{nd} place.

\method\ is model-agnostic and practical for real-world auditing. 
The success of our proposed attack highlights the importance of performing privacy evaluations when sharing synthetic data and motivates the use of defenses such as noise injection, stronger regularization, or differential privacy.
% The success of our proposed attack underscores the need for privacy evaluations alongside synthetic data release and defenses such as noise injection, stronger regularization, or differential privacy.
Looking ahead, we plan to extend \method\ to other generative models (e.g., tabular GANs) and develop defenses that better balance utility and privacy in synthetic data generation.

%% file: content/appendix.tex
% \section*{Appendix}
\appendices
\section{\label{appendix:training}Training and Evaluation Configurations}

This appendix provides the computing infrastructure and software configurations used in our experiments to ensure reproducibility.

All experiments were conducted on a server equipped with an \texttt{NVIDIA RTX 6000 GPU} (48GB VRAM), an \texttt{AMD EPYC 7702P 64-Core Processor}, and 256GB of RAM. The operating system was \texttt{Rocky Linux 9.5 (Blue Onyx)}. All runs used a fixed random seed of 42.

We implemented our code in \texttt{Python 3.10}, and used the following core libraries and frameworks for training and evaluation:
\texttt{torch==2.5.1}, \texttt{catboost==1.2.7}, \texttt{xgboost==2.1.1}, \texttt{scikit-learn==1.4.2}, \texttt{pandas==1.5.3}, \texttt{numpy==1.26.4}, and \texttt{tqdm==4.66.4}.

All models were trained and evaluated using a single GPU. The training pipeline includes feature extraction using attribute prediction models trained on synthetic data, followed by attack classifier training (e.g., CatBoost, XGBoost), and performance evaluation via AUC-ROC and TPR@FPR metrics.